# Healing Data Loss Problems in Android Apps

Oliviero Riganelli, Daniela Micucci, and Leonardo Mariani
*Department of Informatics, Systems and Communication*
*University of Milano Bicocca*
*Viale Sarca 336, IT-20126 Milan, Italy*
Email: {riganelli,micucci,mariani}@disco.unimib.it

*Abstract*—Android apps should be designed to cope with *stop-start events*, which are the events that require stopping and restoring the execution of an app while leaving its state unaltered. These events can be caused by run-time configuration changes, such as a screen rotation, and by context-switches, such as a switch from one app to another. When a stop-start event occurs, Android saves the state of the app, handles the event, and finally restores the saved state. To let Android save and restore the state correctly, apps must provide the appropriate support. Unfortunately, Android developers often implement this support incorrectly, or do not implement it at all. This bad practice makes apps to incorrectly react to stop-start events, thus generating what we defined *data loss problems*, that is Android apps that lose user data, behave unexpectedly, and crash due to program variables that lost their values.

Data loss problems are difficult to detect because they might be observed only when apps are in specific states and with specific inputs. Covering all the possible cases with testing may require a large number of test cases whose execution must be checked manually to discover whether the app under test has been correctly restored after each stop-start event. It is thus important to complement traditional in-house testing activities with mechanisms that can protect apps as soon as a data loss problem occurs in the field.

In this paper we present *DataLossHealer*, a technique for automatically identifying and healing data loss problems in the field as soon as they occur. *DataLossHealer* is a technique that checks at run-time whether states are recovered correctly, and heals the app when needed. *DataLossHealer* can learn from experience, incrementally reducing the overhead that is introduced avoiding to monitor interactions that have been managed correctly by the app in the past.

*Keywords*-Self-healing, Android, Data loss, Stop-start events

## I. INTRODUCTION

When users navigate through, out of, and back to an Android app, they may generate a number of *stop-start events*, which modify the state of the running activity[1]. Handling stop-start events requires putting an app on hold, by saving its internal state, and successively recovering its state to continue with the execution [1], [2]. This process is managed by a set of callback methods implemented by the app. For instance, when the screen is rotated, the Android system invokes the onSaveInstanceState() callback method, destroys the foreground activity, adjusts the orientation, reloads the activity (which is now displayed according to the new orientation), and finally recovers the saved state by invoking the onRestoreInstanceState() callback method.

The default implementation of the callback methods (e.g., the implementation of onSaveInstanceState() and onRestoreInstanceState()) only save and restore the information about the state of the graphical user interface (e.g., the text written in a TextView). Unfortunately, this behavior is often insufficient to correctly handle a stop-start event. Relevant state information can be stored in the member variables of Android activities, and these values are not saved and restored automatically. To handle these cases, Android developers must override the onSaveInstanceState() and the onRestoreInstanceState() methods to save the relevant information and restore it when needed, respectively. Android apps that incorrectly save and restore their state, that is Android apps with data-loss problems, may frequently cause loss of user data, unexpected behaviors, and program crashes.

Data loss problems might be introduced in three ways: by a missing implementation of the callback methods, by incorrectly implemented callback methods, and by upgrades that break the behavior of the callback methods. In the first case, the app incorrectly relies on the default implementation of the callback methods, which omit to save and restore part of the state of the application, such as in the *Mapbox* app where the missing implementation of the callback methods caused the loss of the current map position after every configuration change [?]. In the second case, the app simply overrides the callback methods with a wrong behavior, such as in the *Transmission* app where the incorrect implementation of the callback methods sporadically produced null pointer exceptions [4]. In the third case, the app correctly handles stop-start events, but its behavior is broken by an upgrade of the Android framework, such as in the PagerSlidingTabStrip that can loose the color selected for its ActionBar after the upgrade to API 17 [5]. Introducing any of these problems is relatively easy, as also reported in other studies [6], [7].

Apps that respond unreliably to stop-start events are likely to be annoying for users and fail in the market. It is thus important to implement reliable applications that correctly handle stop-start events and that can be comfortably used in common situations, such as temporary interrupting the activity with an app to answer an incoming call or to retrieve information from another app (e.g., doing a search on Google).

---
[1]Android apps are composed of multiple components called activities.

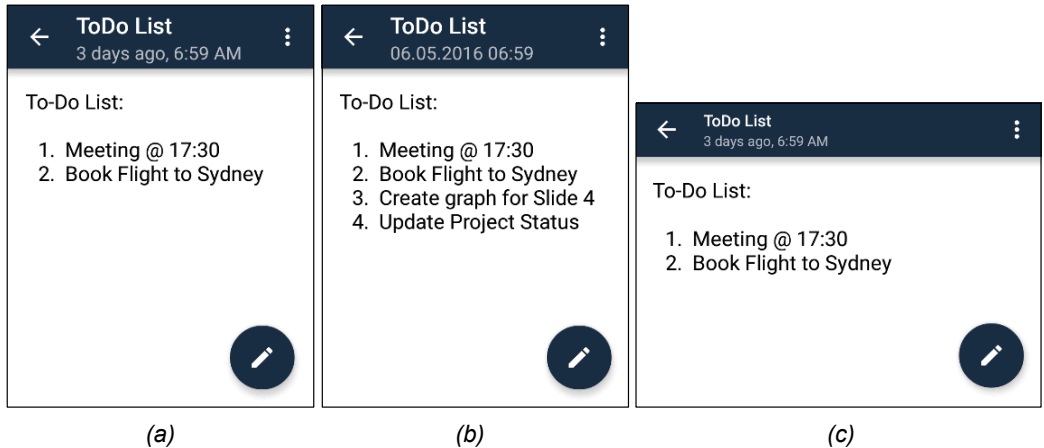

Figure 1: Data loss problem in *ownCloud Notes* app

While there are several testing solutions for Android, such as Robotium [8], Thor [7], and AppDoctor [6], none of these provide specific support to data loss problems detection. Robotium [8] simply offers an environment that testers can exploit to implement capture and replay testing. Thor [7] runs the available test cases while recreating adverse conditions that may expose unexpected problems. AppDoctor [6] can simulate some stop-start events during test execution but its failure detection ability is limited to crashes and any data loss problem that does not cause crashes would remain unnoticed.

Azim et al. address the problem of making apps more reliable by adding the capability to automatically disable the functionalities that caused crashes in the past [9]. This solution suffers from two relevant drawbacks: it progressively reduces the set of functionalities offered by an app and is ineffective with data loss problems that do not cause crashes.

To address the data loss problems that might be experienced in the field by users, we introduce *DataLossHealer*, a self-healing solution for automatically identifying and healing data loss problems in Android apps. *DataLossHealer* is a technique that checks at run-time whether states are recovered correctly after stop-start events, and heals apps when needed. *DataLossHealer* implements the capability to learn from past experience to minimize its intrusiveness and increase its efficiency. In particular, *DataLossHealer* can distinguish the potentially safe from the potentially unsafe situations and activate appropriate countermeasures accordingly.

This paper describes our ongoing effort in the definition of *DataLossHealer* and our experience with a real World application affected by a data loss problem. In the following, we first present a motivating scenario concerning a real data loss problem (Section II), we then illustrate *DataLossHealer* referring to the motivating scenario (Section III), we discuss related work (Section IV), and we finally provide final remarks (Section V).

## II. MOTIVATING EXAMPLE

As motivating example we use an actual Android app affected by a data loss problem. The Android app is *ownCloud Notes* [10], which is an app to create, edit, modify, and delete notes stored on the Cloud. The app is available on the Google Play store.

The app has been affected for a few versions by a data loss problem that can be observed every time a stop-start event is generated while the user is editing a note[2]. Figure 1 shows a sequence of three screenshots that demonstrates how this problem might be experienced in the case of screen rotation: the user first selects an existing note for editing (Figure 1.a shows the note selected for being edited, the note is a ToDo List and includes two tasks), then adds two new tasks to the note (Figure 1.b shows the list containing four tasks), and finally rotates the screen losing all the changes (Figure 1.c shows the list with a wrong date on the header and with two tasks instead of four).

The cause of this faulty behavior is the incorrect implementation of the callback methods managing the stop-start events in the `NoteActivity` class. The data loss problem affects both the date and the content of the ToDo list, which are both restored with outdated values after rotation.

[2]See https://github.com/stefan-niedermann/OwnCloud-Notes/issues/55

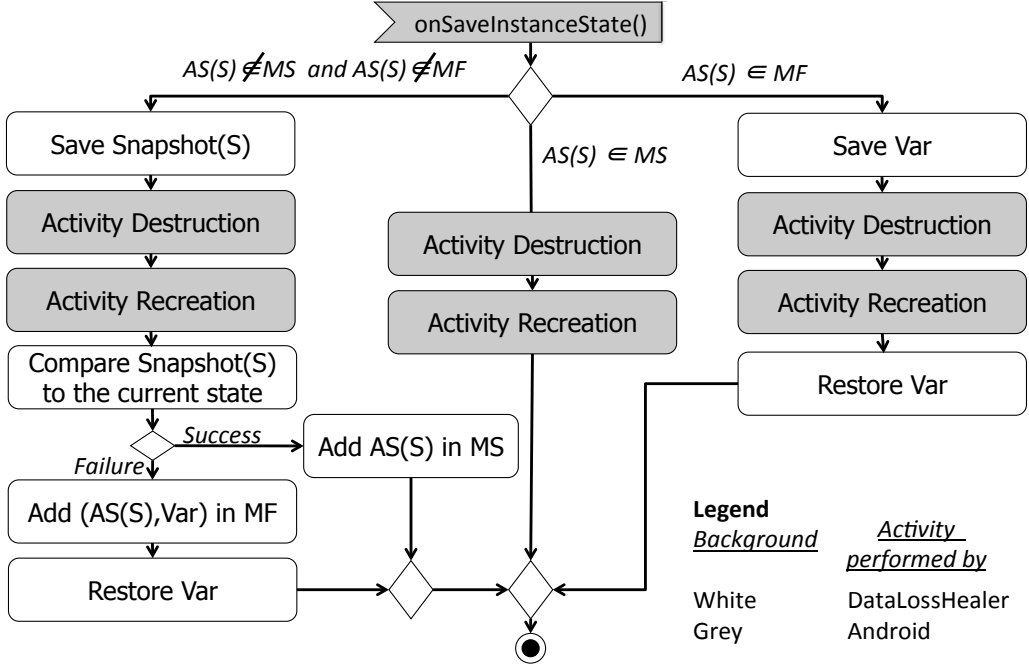

Figure 2: The *DataLossHealer* process.

In the next section, we discuss how *DataLossHealer* can automatically detect and heal this problem.

### III. *DataLossHealer*

The workflow in Fig 2 summarizes the behavior of *DataLossHealer*. The workflow exploits a memory organized in two sets: $MS$ and $MF$. The former set represents the app states where activities have been destroyed and then restored *successfully*, without causing any data loss problem. The latter set represents the app states where the recreation of the activities has *failed* due to a data loss problem. Each entry in $MF$ also includes the list of the variables that lost their values after recreation. Those sets are initially empty and are automatically populated by *DataLossHealer*.

The app states are represented abstractly when entered into $MS$ and $MF$. This is necessary because concrete app states are large, in principle they include the values of every program variable and a representation of the environment, and thus using full state representations may excessively slow down the learning process. Moreover, a same concrete state is unlikely to occur many times in the life of an app, thus the probability to reuse any information learnt in the past would be extremely small. An optimal state abstraction strategy is a strategy that represents in a same way all and only the states that react equivalently to data loss problems. In this way, *DataLossHealer* can simply observe the behavior of the application for one of these states, to generally learn how to handle a broader set of cases.

We now overview how *DataLossHealer* works taking advantage of the workflow shown in Figure 2. When `onSaveInstanceState()` is invoked (see the trigger event in Figure 2), there are three possible ways to continue the execution depending on the value of the abstract representation $AS(S)$ of the current state $S$, as represented by the three main flows in Figure 2:

***New State*** - $AS(S) \notin MS \wedge AS(S) \notin MF$: Since $AS(S)$ does not match with any abstract state observed before ($AS(S)$ is neither in *MS* nor in *MF*), *DataLossHealer* takes a snapshot of the current state $S$ and saves it. The snapshot contains the values of the relevant variables and is later used to check if any data loss problem has occurred (see Section III-C for a more detailed description of the snapshots).

After the activity has been destroyed and recreated, *DataLossHealer* compares the current state of the app to the snapshot persistently saved before. If no data loss is observed, that is the current state matches the snapshot, the abstract state $AS(S)$ is added to *MS*. If a data loss is observed for a set of variables *Var*, the pair $(AS(S), Var)$ is added to *MF* and immediate healing is performed, that is the values in the saved snapshot are used to recover the variable values that have been lost. The next time a stop-start event occurs in the same or a similar state (i.e., in a concrete state that can be mapped to the same abstract state), *DataLossHealer* will handle the case more efficiently exploiting the information recorded in *MS* and *MF*.

**Unsafe State** - $AS(S) \in MF$: Since the values of the variables in *Var* for the pair $(AS(S), Var)$ in $MF$ are likely to get lost once the state of the app has been restored, *DataLossHealer* eliminates the data loss problem by saving the values of the variables in *Var* before the activity is destroyed and then restoring these values after the activity has been recreated. Compared to the *New State* case, *DataLossHealer* efficiently stores only the variables that may get lost, and does not record a whole snapshot of the state.

**Safe State** - $AS(S) \in MS$: Since the state is safe no intervention is required. Compared to the *New State* case, *DataLossHealer* skips taking any snapshot of the system.

Note that the first time a stop-start event occurs in a given state, the *DataLossHealer* performs relatively costly operations, like saving and comparing snapshots. However, *DataLossHealer* learns from past experience and the next time the same (abstract) state occurs, *DataLossHealer* does not take any snapshot but behaves efficiently: if the state is safe, *DataLossHealer* does not intervene at all; while if the state is unsafe, *DataLossHealer* records only the values of the variables that might get lost.

We now describe how we implemented the key operations performed by *DataLossHealer* using the faulty app described in Section II as running example. We consider the case the user rotates the screen after having edited a note.

### A. Monitoring.

*DataLossHealer* needs to intercept the callbacks caused by stop-start events and in particular the callbacks to `onSaveInstanceState()` and `onRestoreInstanceState()` in order to check whether a data-loss problem will occur and, in case, to add the operations necessary to heal the app. There are several frameworks that can be used to this end, such as Xposed [11] and ADBI [12]. Our prototype implementation uses the Xposed Framework [11] because it allows to cost-efficiently intercept method invocations and to change the behavior of an app using run-time hooking and code injection mechanisms.

### B. Generating and Comparing Abstract States.

State information could be abstracted in many ways. We are currently experiencing *DataLossHealer* with a state representation consisting of a pair (*activity*, *BitMask*), where *activity* is the name of the foreground activity, and *BitMask* represents a sequence of binary values with each bit indicating whether each member variable and view of the *activity* has a value different from the default for its type. Since data loss problems cause variables to lose their values after activity recreation, their manifestation intuitively depends on the number of variables that have a non-default value. For a variable assigned with its default value (e.g., a variable of type `Person` assigned to `null`), it is virtually impossible to determine if its value has been lost or not after a stop-start event. Thus, the states with a same set of variables assigned with values different than default should behave the same with respect to their propension to experience any data loss.

Our mask of bits exactly captures this aspect. For instance, in the *ownCloud Notes* app, when the screen is rotated, its abstract state is represented by the pair ⟨*NoteActivity*, *10111111*⟩, which indicates that `NoteActivity` is the foreground activity and that there is only one variable that has a default value (represented by the only zero in the *BitMask*). This variable is the class attribute `notePosition` of type `int` that evaluates to its default value `0` when the screen is rotated. All the other 1s in the mask indicate the presence of activity attributes and views with non-default values.

Trivially, two abstract states $AS_1 = (a_1, bm_1)$ and $AS_2 = (a_2, bm_2)$ are the same if they have the same activity ($a_1 = a_2$) and the same bitmask ($bm_1 = bm_2$).

### C. Recording and Comparing Snapshots

When a state of the app does not correspond to any abstract state in *MS* and *MF* (see left branch in Figure 2), *DataLossHealer* records a snapshot of the current state to be later able to detect data loss problems. Depending on the tradeoff between completeness and cost, the snapshot may include a larger or smaller set of information.

The kind of snapshot we are experiencing with records information about the state of the activity's view hierarchy and the values of the activity's member variables. Intuitively this corresponds in preserving the information shown to the user (the information in the view) and the internal state of the application (the information in the activity).

In the *ownCloud Notes* app, when the `NoteActivity` is the foreground activity and the screen is rotated, the *DataLossHealer* records the values of several variables including the date in the view `mSubtitleTextView` of type `TextView`, the text in the view `noteContent` of type `AppCompatTextView`, and the value of the member variable `note` of type `Note`.

To store a snapshot, *DataLossHealer* reuses the mechanism natively available in Android for saving and loading bundles. A *bundle* is a data structure that is able to persistently store values of certain, mostly primitive, types. The values in the snapshot that are compatible with bundles, such as the member variable `notePosition` of type `int` in the `NoteActivity`, are directly added to the bundle. The incompatible values, such as the custom object `note` of type `Note`, are first converted to the JSON format using the GSON library [13] and then added to the bundle as strings. Our prototype uses this solution to not require anything else than bundles, but other storage technologies might be considered in the future, such as Object Relational Mapping (ORM) frameworks [14] or NoSQL databases [15].

When the app is restored, *DataLossHealer* checks if the views and the member variables are still assigned with values equal to the values in the bundle. Since the activity has been destroyed and then recreated, variables cannot be simply compared by reference. *DataLossHealer* uses the Javers [16] diff framework to compare the state of objects, while it simply directly compares variables of primitive data types and standard view types.

In the *ownCloud Notes* app, after the screen rotation, *DataLossHealer* discovers that a data loss affected three variables: the views `mSubtitleTextView` and `noteContent`, and the member variable `note`. To prevent taking again complete snapshots in similar states in the future, *DataLossHealer* adds the entry ⟨(*NoteActivity*, *10111111*), {`mSubtitleTextView`, `noteContent`, `note`}⟩ to *MF* indicating that the variables `mSubtitleTextView`, `noteContent`, and `note` are the variables to be healed when the foreground activity is *NoteActivity* and its concrete state can be mapped to the abstract state *10111111*.

When no data loss is observed in *ownCloud Notes*, *DataLossHealer* adds an entry to *MS* to prevent taking full snapshots in the future.

In addition to detecting the data loss problem, *DataLossHealer* immediately heals the execution by using the recorded snapshot. The healing procedure simply consists of assigning the saved values to the variables that lost their values. In the *ownCloud Notes* app these variables are `mSubtitleTextView`, `noteContent`, and `note`. Restoring these variables eliminates the data loss problem both visually, because the proper values of the views are restored, and internally to the app, because the proper values of the fields of the activity are restored.

### D. Recording and Restoring Variables

In addition to recording and comparing complete snapshots, *DataLossHealer* records and restores ad-hoc sets of variables based on past experience. In particular, when a stop-start event is detected while an app is in a unsafe state $S$, that is $AS(S) \in MF$, *DataLossHealer* records the values of the variables in *Var* (i.e., the set of variables associated with the data loss for the abstract state $AS(S)$) before the state is destroyed (see right branch in Figure 2).

To save the values of these variables, *DataLossHealer* uses the same mechanism used to take snapshots, that is it uses Android bundles. Note that here *DataLossHealer* efficiently saves only the subset of the state information that might get lost, drastically reducing the amount of information saved. For instance, in the *ownCloud Notes* app, *DataLossHealer* only saves the values of the text in the `mSubtitleTextView` and `noteContent` views, and the value of the member variable `note`, not saving the values of the other views and member variables.

Once the state has been recreated, *DataLossHealer* assigns the values saved in the bundle to the variables subject to data loss. Note that the fix produced by *DataLossHealer* is semantically equivalent to the fix implemented by the developers of the *ownCloud Notes* app.

## IV. RELATED WORK

The development of reliable mobile applications is a complex task with several emerging challenges and issues, such as the unpredictability and the high variability of the application context [17], [18], the instability and the rapid evolution of the environment [19], [20], and the lack of robust and reliable testing and analysis tools [21], [22]. Data loss problems are popular problems [6], [7] caused by the difficulty of properly implementing the interaction between an app and both the underlying framework and its environment, as well as the difficulty of predicting every possible execution scenario.

In this paper we presented a self-healing approach to prevent the occurrence of any data loss directly in the field while the app is running. There are two main classes of approaches related to our contribution: testing techniques for Android applications and self-healing techniques for Android applications.

### A. Testing Android applications

To a significant extent, the validation of mobile applications is still a manual activity [21], supported by products that can only partially automate the testing process, such as Robutium [8] that can automate the execution of test cases that have been already executed manually.

In the last few years the research community has been particularly active in the definition of solutions for increasing the level of automation of validation techniques for Android applications, and mobile applications in general. Most of the results concern techniques addressing specific situations, such as the validation of exceptional behaviors [7], [23], the discovery of security and privacy concerns [24], [25] and the detection of energy and performance bugs [26], [27].

Although all these techniques might be useful to reveal a range of bugs, none of them specifically addresses data loss problems. To the best of our knowledge, AppDoctor [6] is the only technique that can generate test cases that include stop-start events. However, AppDoctor is equipped with an oracle that can only detect crashes, thus the many data loss problems that cause loss of data without producing crashes would remain undetected if addressed with AppDoctor (consider for instance the data loss problem presented in Section II). Moreover, AppDoctor can only simulate some of the many stop-start events that can be produced while using an app, thus covering only a small subset of the situations that might be experienced in the field.

In order to protect apps from the unexpected behaviors that might be experienced as consequence of a data loss problem, it is necessary to extend their capability with the ability to prevent and overcome data loss at run-time. *DataLossHealer* represents our first attempt to design a solution going into this direction.

### B. Healing Android applications

Self-healing techniques have been extensively studied as solution to overcome failures in many different contexts, including Web Applications [28], Operative systems [29], and the Cloud [30], while the domain of mobile applications has surprisingly not been well explored yet, even if there are many classes of environment and context dependent problems that would benefit from self-healing solutions.

A few techniques investigated how to dynamically inject patches into a target application to fix vulnerabilities in Android applications [31], [32]. While these techniques present interesting solutions to dynamically modify the behavior of Android applications, they do not represent actual self-healing solutions because the code that implements the patch is produced offline and then only deployed online. On the contrary, *DataLossHealer* analyzes applications and takes decisions online, without any input required from users or developers.

In a recent work, Azim et al. [9] proposed a self-healing solution that can detect crashes in Android apps and avoids future occurrences of the same crashes by bypassing the execution of the code that caused the crash. The approach focuses on failures caused by unhandled exceptions, while the healing actions consist of adding try/catch blocks that capture the unhandled exceptions and skip the execution of the problematic activities.

Different from this work, *DataLossHealer* focuses on a specific problem, that is data loss problems, rather than general program crashes, with the advantage of implementing a mechanism to completely heal failing executions, without impacting on the functionalities offered by the application, instead of simply skipping the execution of some functionalities without healing them.

## V. CONCLUSION

While using mobile devices, users generate stop-start events, which are events that cause the foreground application to be stopped and successively restored. Frequent cases where these events are generated include receiving an incoming call, switching from one app to another, and rotating the screen of the smartphone.

In all these cases, the state of the app must be properly saved and restored, otherwise data might be lost resulting in a data loss problem. Data loss problems might be particularly annoying for the users, who might experience unexpected behaviors and crashes, in addition to losing their data.

Since developers have to implement an appropriate support to stop-start events in their apps, and many times this support is implemented incorrectly or not implemented at all, data loss problems are also common [6], [7].

In this paper, we propose to address this problem with the *DataLossHealer* self-healing solution, which can detect and heal executions that cause data loss problems. *DataLossHealer* works by saving the state of the apps before and after a stop-start event occurs, to automatically detect data-loss problems. Interestingly, *DataLossHealer* can increase its efficiency over time by learning from experience. The learning task is aimed to minimize the frequency of intervention, preventing to record state information when the app is in states known to be safe, and to reduce the cost of the operations, preventing to save information about variables that are known to be unaffected by the data loss.

We reported our experience with the *ownCloud Notes* app [10], which is an app affected by an actual data loss problem. In the future, we plan to experience our solution with a range of Android apps to extensively evaluate its effectiveness, as well as investigating alternative ways of representing the abstract states, and mechanisms to save and restore state information.

### ACKNOWLEDGMENT

This work has been partially supported by the H2020 Learn project, which has been funded under the ERC Consolidator Grant 2014 program (ERC Grant Agreement n. 646867).